\documentclass{ws-p9-75x6-50}

\def\aap{A\&A}
\def\aaps{A\&A Supp.}
\def\aj{AJ}
\def\apj{ApJ}
\def\apjl{ApJ Lett.}
\def\nat{Nature} 
\def\mnras{MNRAS}

\begin{document}

\title{Radio Plasma as a Cosmological Probe}

\author{Torsten A.  En{\ss}lin} 

\address{Max-Planck-Institut f\"{u}r Astrophysik,
 Karl-Schwarzschild-Str.1, 85740 Garching, Germany}

\maketitle

\abstracts{
Plasma containing relativistic particles appears in various
forms in the inter galactic medium (IGM): As radio plasma released by
active radio galaxies, as fossil radio plasma from former radio
galaxies -- so called {\it radio ghosts}--, as {\it cluster radio
relics} in some clusters of galaxies, and as {\it cluster radio
halos}. The impact of the different forms of radio plasma on the IGM
and their use as diagnostic tools are briefly discussed.
}

\section{Radio Galaxies}
Radio galaxies produce radio cocoons, which create large
cavities\cite{2000MNRAS.318L..65F} in the IGM gas filled with magnetic
fields and relativistic particles: electrons and positrons or
protons. The electrons (and positrons if present) of several GeV
reveal their presence by synchrotron emission at radio
wavelength. From the observed radio emission the minimal energy
density in the radio plasma can be deduced. Since the radio plasma has
to have a higher (or equal) pressure than the environment, a direct
probe of the ambient gas pressure is
given\cite{1992A&A...265....9F}. Also the ram pressure effects of
swept up IGM material during the expansion of the cocoon can be used
to determine IGM gas
densities\cite{1998A&A...329..431M,2000A&AS..146..293S}. The
morphology of the radio galaxy can give information about relative
motions of the galaxy and the IGM\cite{1998Sci...280..400B}, but also
allows to detect IGM shock waves\cite{ngc315}.

The amount of energy released by radio galaxies into the gaseous
Universe during the whole history of the Universe is large. Rough
estimates give a ratio of 0.1 ... 1 between the energy released by
radio galaxies and by the gravitationally driven structure
formation\cite{1998A&A...333L..47E,2000IAUJD..10E...9E}. Radio
galaxies are therefore one of the best candidates in order to explain
the necessary non-gravitational heating required by the observed
entropy-floor of clusters and groups of
galaxies\cite{1999Natur.397..135P,2000MNRAS.318..889W}.  Therefore remnant radio plasma
should be a ubiquitous phase of the IGM.

\section{Radio Ghosts}

Radiative energy losses let the radio emitting electrons in radio
cocoons become invisible to our instruments within cosmologically
short times ($10^8$ year). Afterwards, the cocoon of fossil radio
plasma is named {\it radio ghosts}\cite{Ringberg99}.  But the
electrons confined by the magnetic fields should still stay
relativistic, and might reveal their presence by inverse Compton
effects on the cosmic microwave background (CMB): the relativistic
Sunyaev-Zeldovich (rSZ) effect. Since a CMB photon scattered by a
relativistic electron gains a large factor in energy, it is
practically removed from the CMB spectrum. Thus the rSZ effect is
mostly an absorption at CMB frequencies. The optical depth of the
total electron population residing in radio ghosts is of the order of
$\tau \sim 10^{-7}$, but this number has large
uncertainties\cite{2000A&A...360..417E}. Since the expansion of the
radio cocoons into the IGM should at least release some pressure work
to the environmental gas, the corresponding thermal SZ effect should
have a Comptonization parameter of the
order\cite{yamada1999,2000A&A...360..417E} of $y \sim 10^{-5...-6}$.

But not only the relativistic electrons may reveal their presence by
scattering of radiation. Also the magnetic fields of radio ghosts are
able to deflect charged particles even at the highest energies
observed in the cosmic ray spectrum. If the spatial distribution of
radio ghosts is sufficiently unclustered compared to the clustering of
galaxies, then the expected number densities of ghosts would be
sufficient to explain the mysterious isotropy of the ultra high energy
cosmic ray particles, even if the sources are as inhomogeneously
distributed as the galaxies\cite{MedinaTanco2000}.

\section{Cluster Radio Relics}

Recently, radio ghosts were detected by X-ray deficits in cluster core
regions\cite{2000MNRAS.318L..65F}. But very likely, they already
showed up as {\it cluster radio relics} 20 years ago. These
extended radio sources in some clusters of galaxies are not specially
connected to any optical galaxy\cite{1996IAUS..175..333F}. Cluster
radio relics are preferentially located in peripheral regions of
clusters. They have usually steep radio spectra and exhibit often a
high degree of linear polarization. In several cases, their locations
can be associated with strong shock waves in merging clusters of
galaxies\cite{1998AA...332..395E}. In the cases of Abell 2256 and
Abell 1367 temperature substructures of the hot ICM gas could be
detected\cite{1994Natur.372..439B,1998ApJ...500..138D}, which support
the presence of a shock wave at the location of cluster relics in
these clusters. For Abell 754\cite{1998ApJ...493...62R,kassim2001a},
Abell 2256\cite{1995ApJ...453..634R}, Abell
3667\cite{1999ApJ...518..603R} and also the
Coma\cite{1994ApJ...427L..87B} cluster numerical simulations of merger
events were satisfactorily fitted to the X-ray data, which also
supports the shock wave-relic connection. The mechanism producing the
cluster radio relics is very likely adiabatic compression of radio
ghosts in an environmental shock wave\cite{ensslin2000b}. This
strengthens the internal magnetic fields and shifts the electron
population to higher energies. If the upper cooling cutoff of the
electron spectrum can be shifted above the radio emitting energies,
the ghost's radio emission is revived and the ghost appears as a
cluster radio relic. Cluster radio relics therefore not only allow the
study of shock waves, but also give a view into the fossil
Universe. Their number should be much higher at lower frequencies, due
to the distribution of frequency cutoffs in their radio spectra.

\section{Cluster Radio Halos}

Cluster radio halos appear preferentially in the center of clusters of
merging galaxies\cite{1996IAUS..175..333F,1999NewA....4..141G}. They
have morphologies similar to the X-ray morphologies of their host
clusters\cite{govoni2000sub}. Radio polarization could not be reported
in any case. Their large physical size ($\sim$ Mpc) require that the
radio emitting electrons are accelerated or injected throughout the
cluster volume. 

The energy source of all non-thermal processes in clusters should be
either the kinetic energy of matter falling onto clusters, or the
outflows from galaxies. The latter can be divided in galactic winds,
which are strongest for starburst galaxies, and ejection of radio
plasma from an AGN. All these processes can produce shock waves and
inject turbulence into the ICM, and therefore produce conditions where
Fermi mechanisms accelerate particles. For a brief review see
En{\ss}lin (1999)\cite{Pune99}.

A promising injection mechanism of relativistic electrons is secondary
particle production from hadronic interactions of relativistic protons
with the background
gas\cite{1980ApJ...239L..93D,1982AJ.....87.1266V,1999APh....12..169B,2000A&A...362..151D}:
\begin{eqnarray}
  \label{eq:pp}\nonumber
  p + p &\rightarrow& 2 \,N + \pi^\pm \\
  \pi^\pm &\rightarrow& \mu^\pm + \nu_{\mu}/\bar{\nu}_{\mu} \rightarrow
  e^\pm + \nu_{e}/\bar{\nu}_{e} + \nu_{\mu} + \bar{\nu}_{\mu}\nonumber
\end{eqnarray}
The lifetime of relativistic protons in the ICM is of the order of the
Hubble time, or
larger\cite{1997ApJ...477..560E,1997ApJ...487..529B}. Thus they are
able to travel large distances from their sources before they release
their energy. The production of electrons via charged pions has to be
accompanied by gamma ray production via neutral pions 
\cite{1982AJ.....87.1266V,1997ApJ...487..529B,1998APh.....9..227C,2000A&A...362..151D}:
\begin{eqnarray}
  \label{eq:ppgamma}
  p + p &\rightarrow& 2\, N + \pi^o \nonumber\\
  \pi^o &\rightarrow& 2 \, \gamma \nonumber
\end{eqnarray}
Thus clusters with radio halos might have gamma-ray halos, which would
be, if detected, a direct proof for a hadronic origin of radio halos.

\section{Conclusions}

{\bf Radio galaxies:}
\begin{itemize}
\item
probe the thermodynamical state of the gaseous Universe.
\item 
produce large amounts of fossil radio plasma.
\end{itemize}
{\bf Radio ghosts}, the invisible descendents of radio galaxies:
\begin{itemize}
\item
may produce an relativistic Sunyaev-Zeldovich effect.
\item 
scatter and possibly isotropize ultra high energy cosmic ray particles.
\end{itemize}
{\bf Cluster radio relics:}
\begin{itemize}
\item
trace shock waves of the large scale structure formation flows.
\item 
are likely revived radio ghosts.
\end{itemize}
{\bf Cluster radio halos:}
\begin{itemize}
\item
indicate recent cluster merger events.
\item 
trace non-thermal processes in the intracluster medium.
\end{itemize}


\begin{thebibliography}{10}

\bibitem{2000MNRAS.318L..65F}
A.~C. {Fabian}, J.~S. {Sanders}, S.~{Ettori}, G.~B. {Taylor}, S.~W.
  {Allen}, C.~S. {Crawford}, K.~{Iwasawa}, R.~M. {Johnstone}, and P.~M.
  {Ogle}.
 {\em \mnras}, 318, L65, 2000.

\bibitem{1992A&A...265....9F}
L.~{Feretti}, G.~C. {Perola}, and R.~{Fanti}.
 {\em \aap}, 265, 9, 1992.

\bibitem{1998A&A...329..431M} K.-H. Mack, , U. Klein, 
C.~P. O'Dea,  A.~G. Willis, and L. Saripalli, {\em  \aap}, 329, 431, 1998. 

\bibitem{2000A&AS..146..293S} A.~P. Schoenmakers, 
K.-H. Mack, A.~G. de Bruyn,  H.~J.~A. R{\"o}ttgering,U.  Klein, and
H. van der Laan, {\em  \aaps}, 146, 293, 2000.




\bibitem{1998Sci...280..400B}
J.~O. {Burns}.
 {\em Science}, 280, 400, 1998.


\bibitem{ngc315}
T.~A. En{\ss}lin, P. Simon, P.~L. Biermann, U. Klein, S. Kohle,
P.~P. Kronberg, and K.-H. Mack. {\em \apjl}, in press, 
astro-ph/0012404

\bibitem{1998A&A...333L..47E} T.~A. Ensslin, Y. Wang, B.~B. Nath, and
P.~L. Biermann, {\em \aap}, 333, L47, 1998.


\bibitem{2000IAUJD..10E...9E} T.~A. En{\ss}lin, Cluster Mergers and
their Connection to Radio Sources, 24th meeting of the IAU, Joint
Discussion 10, Manchester, England., 10, E9, 2000. astro-ph/0011052


\bibitem{1999Natur.397..135P}
T.~J. {Ponman}, D.~B. {Cannon}, and J.~F. {Navarro}.
 {\em \nat}, 397, 135, 1999.

\bibitem{2000MNRAS.318..889W} K.~K.~S. Wu, 
A.~C. Fabian, and P.~E.~J. Nulsen, {\em  \mnras}, 318, 889, 2000.

\bibitem{Ringberg99} T.~A. {En{\ss}lin}. In {\em Ringberg Workshop on
`Diffuse Thermal and Relativistic Plasma in Galaxy Clusters'}
eds.{H. B{\"o}hringer, L. Feretti, P.  Schuecker}, MPE Report, 271,
275, 1999., astro-ph/9906212


\bibitem{2000A&A...360..417E}
T.~A. {En{\ss}lin} and C.~R. {Kaiser}.
 {\em \aap}, 360, 417,  2000.

\bibitem{yamada1999}
M. Yamada, N. Sugiyama, and J. Silk, ApJ 522, 66, 1999.

\bibitem{MedinaTanco2000}
G.~{Medina-Tanco} and T.~A. {En{\ss}lin}.
 {\em Astroparticle Physics}, in press, 2000.

\bibitem{1996IAUS..175..333F}
L.~{Feretti} and G.~{Giovannini}.
 In {\em IAU Symp. 175,  Extragalactic Radio Sources}, 
  333, 1996.

\bibitem{1998AA...332..395E}
T.~A. {En{\ss}lin}, P.~L. {Biermann}, U. {Klein}, and S. {Kohle}.
 {\em \aap}, 332, 395,  1998.

\bibitem{1994Natur.372..439B}
U.~G. {Briel} and J.~P. {Henry}.
 {\em \nat}, 372, 439,  1994.

\bibitem{1998ApJ...500..138D}
R.~H. {Donnelly}, M.~{Markevitch}, W.~{Forman}, C.~{Jones}, L.~P. {David},
  E.~{Churazov}, and M.~{Gilfanov}.
 {\em \apj}, 500, 138, 1998.

\bibitem{1998ApJ...493...62R}
K.~{Roettiger}, J.~M. {Stone}, and R.~F. {Mushotzky}.
 {\em \apj}, 493, 62, 1998.

\bibitem{kassim2001a}
N.~{Kassim}, T.~E. {Clarke}, T.~A. {En{\ss}lin}, A.~S. {Cohen}, and
  D.~{Neumann}.
 {\em \apjl}, submitted, 2001.

\bibitem{1995ApJ...453..634R}
K.~{Roettiger}, J.~O. {Burns}, and J.~{Pinkney}.
 {\em \apj}, 453, 634, 1995.

\bibitem{1999ApJ...518..603R}
K.~{Roettiger}, J.~O. {Burns}, and J.~M. {Stone}.
 {\em \apj}, 518, 603, 1999.

\bibitem{1994ApJ...427L..87B}
J.~O. {Burns}, K.~{Roettiger}, M.~{Ledlow}, and A.~{Klypin}.
 {\em \apjl}, 427, L87, 1994.

\bibitem{ensslin2000b}
T.~A. {En{\ss}lin} and Gopal-{Krishna}.
 {\em \aap}, in press, 2001.

\bibitem{1999NewA....4..141G}
G.~{Giovannini}, M.~{Tordi}, and L.~{Feretti}.
 {\em New Astronomy}, 4, 141, 1999.

\bibitem{govoni2000sub}
F.~{Govoni}, T.~A. {En{\ss}lin}, L.~{Feretti}, and G.~{Giovannini}.
 {\em \aap}, submitted, 2000.

\bibitem{Pune99} T.~A. {En{\ss}lin}.  In {\em IAU Symp. 199,  `The
Universe at Low Radio Frequencies'}, 1999.  astro-ph/0001433.

\bibitem{1980ApJ...239L..93D}
B.~{Dennison}.
 {\em \apjl}, 239, L93, 1980.

\bibitem{1982AJ.....87.1266V}
W.~T. Vestrand.
 {\em \aj}, 87, 1266, 1982.

\bibitem{1999APh....12..169B}
P.~{Blasi} and S.~{Colafrancesco}.
 {\em Astroparticle Physics}, 12, 169, 1999.

\bibitem{2000A&A...362..151D}
K.~{Dolag} and T.~A. {En{\ss}lin}.
 {\em \aap}, 362, 151, 2000.

\bibitem{1997ApJ...477..560E}
T.~A. {En{\ss}lin}, P.~L. {Biermann}, P.~P. {Kronberg}, and
  X.-P. {Wu}.
 {\em \apj}, 477, 560, 1997.

\bibitem{1997ApJ...487..529B}
V.~S. {Berezinsky}, P.~{Blasi}, and V.~S. {Ptuskin}.
 {\em \apj}, 487, 529, 1997.

\bibitem{1998APh.....9..227C}
S.~{Colafrancesco} and P.~{Blasi}.
 {\em Astroparticle Physics}, 9, 227, 1998.

\end{thebibliography}

\end{document}